\documentclass[epjCONF,columns,a4paper,indent]{svjour} 
\usepackage[left=2.5cm,right=1.5cm,bottom=2.5cm,top=4.5cm]{geometry}

\usepackage{graphics}
\usepackage[varg]{txfonts} 
\usepackage[latin1]{inputenc}
\session-title{Hadron Collider Physics Symposium 2011, November 14-18, Paris, France}

\begin{document}
\title{Heavy flavour production and spectroscopy at LHCb\thanks{Presented at Hadron Collider Physics Symposium 2011, November 14-18, Paris, France}}
\author{Diego Milan\'es\thanks{\email{diego.milanes@ba.infn.it}}, on behalf of the LHCb Collaboration }
\institute{INFN, Sezione di Bari}
\abstract{
We summarize the main measurements performed with the LHCb detector on production and spectroscopy in the heavy flavour sector, using data samples recorded during 2010 and 2011 data taking in proton-proton collisions at $\sqrt s=7$\,TeV .
} 
\maketitle
\section{Introduction}
\label{intro}
Measurements of the heavy quark production cross-sections in proton-proton collisions test the predictions of quantum chromodynamics (QCD), and probes the spectrum and dynamics of the colliding partons. The formation of heavy quarkonia states (bounded $q\bar q$) can be factorized into two steps according to quantum chromodynamics (QCD). First, the creation of a $q\bar q$ pair via small-distance interactions (perturbative), followed by the evolution into a quarkonium state via the exchange of soft gluons (non perturbative). The non-relativistic QCD (NRQCD), predicts the probability of a heavy $q\bar q$ pair to evolve into heavy quarkonium, as a function of color-singlet (CS) \cite{CS} and color-octet (CO) \cite{CO} matrix elements. The CS model, provides a leading-order (LO) description of quarkonium formation, but it underestimates the measured cross-section for $J/\psi$ production at Tevatron~\cite{csteva}, particularly at high transverse momentum $p_T$. This discrepancy can be reduced including processes such as quark and gluon fragmentation, but still it fails to reproduce the measured cross-section. The introduction of CO model, with the matrix elements tuned to data, improves the description of the measured shape and magnitude of the $J/\psi$ cross-section. Recent theoretical studies~\cite{th-onia} incorporate higher order corrections to the CS models, reducing significantly the observed discrepancy in quarkonium production without including CO matrix elements. However, the agreement is still not perfect, leaving open the question of a complete description of quarkonia formation \cite{quarkonia}.

The spectrum of heavy hadrons has been predicted using QCD potentials and chiral models \cite{spec-theo}. Discrepancies between predictions and observations, mainly for high mass states \cite{spec-exp}, makes spectroscopy an active and often controversial field.

\section{The LHCb detector} 
\label{detector}
The LHCb detector is a single-arm forward spectrometer covering a unique rapidity range ($2<\eta<5$), where $b\bar b$ cross section is peaked. Due to this, although it covers $\sim$4\% of the solid angle, it detects $\sim$40\% of heavy quark production cross-section.
The LHCb detector \cite{lhcb} is specialized in beauty and charm physics studies, exhibiting outstanding tracking, vertexing and particle identification capabilities. In 2010 and 2011, the detector recorded about 1.1\,fb$^{-1}$ of integrated luminosity in proton-proton collisions at $\sqrt s =$7\,TeV, more than the 90\% of the luminosity delivered by the Large Hadron Collider (LHC).

\section{Heavy flavour production}
\label{production} 
In a general form, the production cross-section at proton-proton collisions of a given state $A$ can be written as
\begin{equation}
\sigma(pp\to A)=\frac{N(A\to f)}{\mathcal{L}\cdot\epsilon\cdot\mathcal{B}(A\to f)},
\end{equation} 
where $N(A\to f)$ is the number of observed signal events of $A$ decaying into the final state $f$, $\mathcal{L}$ is the integrated luminosity of the sample, $\epsilon$ is the efficiency which accounts for trigger and reconstruction effects, and $\mathcal{B}(A\to f)$ is the branching fraction of the decay.

 Using the 2010 data sample of 40\,pb$^{-1}$, the LHCb collaboration has measured several production cross-sections such us open charm \cite{opencharm}, $J/\psi$ \cite{jpsi}, double $J/\psi$ \cite{doublejpsi}, $\psi(2S)$ \cite{psi2s}, $\chi_{c2}/\chi_{c1}$ ratio \cite{chic}, $B^\pm$ \cite{B} and $\Upsilon(1S)$ \cite{upsilon1s}. Due to limited space we do not discuss all the analyses in this document.

\subsection{Double $J/\psi$ production}
Theoretical calculations based on LO production of CS-states predict a total cross-section of 24\,nb for $J/\psi J/\psi$ production in the current LHC running conditions \cite{jpsijpsith}. This calculation accounts for additional feed-down from $J/\psi \psi(2S)$ and $\psi(2S)\psi(2S)$ production, but not for double parton scattering. The extrapolated prediction to the angular LHCb coverage gives a cross-section of about 4\,nb with a 30\% of uncertainty.

For this analysis, we use an integrated luminosity of 37.5\,pb$^{-1}$, collected by the LHCb detector between July and November of 2010. $J/\psi$ meson candidates are reconstructed from a pair of oppositely-charged tracks identified as muons. Therefore, we select events with 4 of these tracks, originated from a common vertex, with good track quality and with $p_T<650$\,MeV/$c$. Muon identification is achieved by comparing a global likelihood function provided by the particle identification, tracking and calorimeter subdetectors, with the corresponding likelihood for light hadrons. Selected $\mu^+\mu^-$ candidates, with an invariant mass of 3.0-3.2\,GeV$/c^2$, are paired to form $(\mu^+\mu^-)_1(\mu^+\mu^-)_2$ combinations. Background coming from $J/\psi$ mesons associated to different production vertex and from $B$ decays is removed by applying quality requirements to the vertex fit, constraining the 4 muons to come from the same vertex and requiring the vertex to be compatible with one of the primary vertices produced after the proton-proton collision.

\begin{figure}
\resizebox{1.0\columnwidth}{!}{\includegraphics{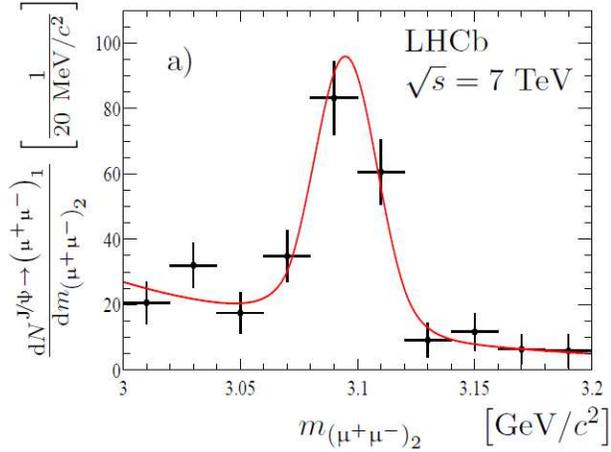} }
\caption{Raw signal yield of $J/\psi\to(\mu^+\mu^-)_1$ in bins of $(\mu^+\mu^-)_2$ invariant mass, as observed in data.}
\label{fig:1}      
\end{figure}

The number of events with two $J/\psi$  mesons is extracted from the single $J/\psi$ mass spectrum. The invariant mass distributions of the first muon pair are obtained in bins of the invariant mass of the second pair, where the first $\mu^+\mu^-$ pair is chosen to be the one with smaller $p_T$. The signal is described using a Crystal-Ball function plus an exponential to describe the background component. The position of the $J/\psi$ peak is extracted from an inclusive $J/\psi$ sample. The fit result projected on the data sample is shown in Fig.~\ref{fig:1}, where the extracted yield is $N(J/\psi J/\psi)=141\pm19$ events. The yield of events with both $J/\psi$ mesons in the detector fiducial range and explicitly triggered by one $J/\psi$ through the dedicated muon trigger lines is found to be $116\pm16$. This is the sample used to extract the cross-section.

The efficiency evaluation accounts for reconstruction, muon misidentification and trigger effects. The efficiency is factorized as the product of the efficiency for each $J/\psi$ meson. The reconstruction efficiency $\epsilon_R=\epsilon_{R1}\times\epsilon_{R2}$, is the product of the efficiency of the two $J/\psi$ mesons, and it is a function of the geometrical acceptance variables $p_T$ and $\eta$ and $|\cos\theta^*|$, where $\theta^*$ is the angle between the $\mu^+$ momentum in the $J/\psi$ rest frame and the $J/\psi$ momentum in the laboratory frame. The angular dependence due to the unknown polarization of the $J/\psi$ candidate is computed from a sample of $J/\psi\to\mu^+\mu^-$ simulated events. Muon identification efficiency is extracted from the analysis of the inclusive $J/\psi$ sample. 
The trigger efficiency is determined with a sample of independently triggered events. 
The effect of the global event cuts applied in the trigger has been studied in detail for inclusive $J/\psi$ events, and applied here assuming factorization.

The largest systematic contribution is due to the knowledge of the track-finding efficiency. A 4\% uncertainty per track is assigned, based on studies comparing the reconstruction efficiency in data and simulation, using a tag-and-probe approach.  Additional systematic uncertainties in the method to extract the efficiency have been accounted for, as well as data-simulation discrepancies. 
The luminosity was determined in specific periods during data taking, using both Van der Meer scans and beam gas imaging method \cite{lumi}. The systematic uncertainty associated with these methods is of 10\%. The total systematic contribution to the cross-section uncertainty was found to be of 21\%.

\begin{figure}
\resizebox{1.0\columnwidth}{!}{\includegraphics{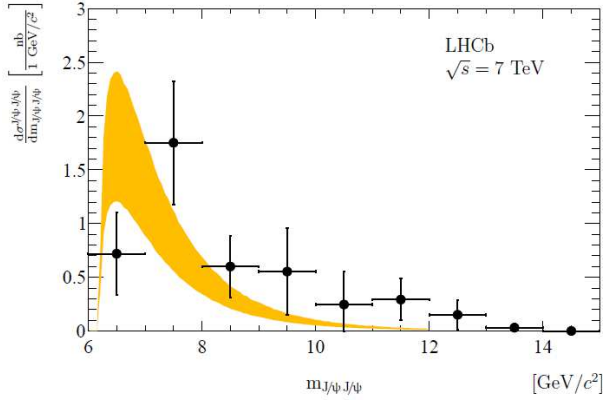} }
\caption{Differential production cross-section for $J/\psi$  pairs as a function of the invariant mass of the $J/\psi J/\psi$ system. The points correspond to the data. Only statistical uncertainties are included in the error bars. The shaded area corresponds to prediction by the model described in \cite{jpsijpsith}.}
\label{fig:2}      
\end{figure}

Using the value of $\mathcal{B}(J/\psi\to\mu^+\mu^-)=(5.93\pm0.06)\%$ \cite{pdg}, the cross-section result is $\sigma(J/\psi J/\psi) = 5.1\pm1.0\pm1.1$\,nb, where the first uncertainty is statistical and the second systematic. The differential production cross-section of $J/\psi$ pairs as a function of the invariant mass
of the $J/\psi J/\psi$ system is shown in Fig.~\ref{fig:2}, where we observe a reasonable agreement, within uncertainties, between the measurement performed and the prediction.

\section{Heavy flavour spectroscopy}
\label{spectroscopy} 
Spectroscopy is one of the many fronts in which the LHCb collaboration is actively working. To improve our knowledge and understanding on how quarks interact among them and recombine them with others to form intriguing states, it is necessary to study the spectra of the different families of mesons and baryons, looking for unknown objects and determining their properties. 

We will discuss studies on production of $D_{(s)J}$ and $B_{(s)J}$ \cite{bst} states, as well as searches of these states decaying into $Ah$ final states, with $h=\pi^\pm$, $K^\pm$ or $K_S^0$, and $A=D^0$, $D^+$, $D^{*+}$, $B^0$ or $B^+$, aimed to confirm the presence of high mass structures observed in other experiments \cite{djdsj,bstneutral}, and explore the higher mass part of the spectra. In addition, mass measurements of the exotic $X(3872)$ meson \cite{x3872} and heavy baryons such as $\Omega_b^-$ and $\Xi_b^-$ \cite{barmass}, will be performed.\\

\subsection{Orbitally excited $B^{**}$ mesons}
\label{bst} 
The properties of the excited $B$ mesons containing a light quark ($B^+$, $B^0$, $B^0_s$ ) are predicted by Heavy Quark Effective Theory (HQET) in the limit of infinite $b$-quark mass \cite{spec-theo,bstth}. Under the heavy quark approximation the $B$ mesons are characterized by three quantum numbers: the orbital angular momentum $L$ ($S$, $P$, $D$ for L = 0, 1, 2 respectively), the angular momentum of the light quark $j_q=|L\pm1/2|$ , and the total angular momentum $J = |j_q \pm1/2 |$. For $L = 1$ there are four different possible $(J; j_q)$ combinations, all parity-even. These are known as the orbitally excited states or $B^{**}$ states. Among these states we have the $B_1(5721)^0$ and $B_2^*(5747)$, observed in $B^{*+}\pi^-$ and $B^+\pi^-$ decays \cite{bstneutral}. At LHCb, we reconstruct these decay modes, but also the $B^0\pi^+$ and $B^{*0}\pi^+$ where we must observe the isospin partners of the mentioned states.

In this analysis, we use an integrated luminosity of 336.5\,pb$^{-1}$, collected by the LHCb detector between May and July of 2011. The soft photons from the $B^{*0}$ decay are not reconstructed, therefore objects decaying to both $B^{*0}\pi^+$ and $B^0\pi^+$ are expected to show two peaks in the $B^0\pi^+$ invariant mass distribution, separated by 
a quantity corresponding to the $M(B^{*0})-M(B^0)$ mass difference. The $B^0$ meson is reconstructed into the following final states: $J/\psi(\mu^+\mu^-)K^*(892)^0(K^+\pi^-)$, $D^-\pi^+$ and $D^-\pi^+\pi^+\pi^-$, with $D^-\to K^+\pi^-\pi^-$. We combine the $B^0$ candidate with tracks, which are required to originate from the same proton-proton interaction. The companion track is required to be identified as a pion, to have good quality track fit, $p_T>1$\,GeV/$c$ and $p>10$\,GeV/$c$. For convenience, we study the invariant mass relative to the threshold, which has the form $Q(B^0\pi^+) = M(B^0\pi^+)-M(B^0)-M(\pi^+)$, where $M(B^0)$ and $M(\pi^+)$ are the nominal masses of the mesons quoted by Ref.~\cite{pdg}.

\begin{figure}
\resizebox{1.0\columnwidth}{!}{\includegraphics{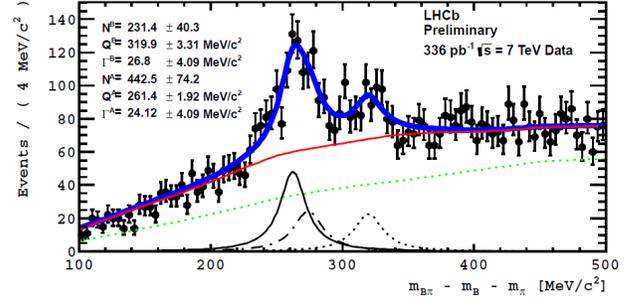} }
\caption{$Q(B^0\pi^+)$ distribution. LHCb data (points) and the total fit (blue) are superimposed. The components of the PDF are, combinatorial background and associated production (red), combinatorial background (dashed green), $B_1^+\to B^{*0}\pi^+$ (solid black), $B_2^{*+}\to B^{*0}\pi^+$ (dot-dashed black) and $B_2^{*+}\to B^0\pi^+$ (dotted black).}
\label{fig:3}      
\end{figure}

Combinatorial background shape is extracted from data, using a sample of reconstructed $B^+\pi^+$ combinations, since the wrong-sign decay $B^0\pi^-$ has structures created by contributions from $B^0-\bar B^0$ mixing. A significant excess, not attributed to any resonant state, is observed in the $Q(B^0\pi^+)$ with respect to the $Q(B^+\pi^+)$ distribution, interpreted as an associated production due to $b$-jet hadronization. This component is described using a kernel-like distribution. Signal resonances in the $Q$-distribution are modeled using relativistic Breit-Wigner lineshapes. The detector resolution is about 3\,MeV/$c^2$ and can safely be neglected. The fit results are shown in Fig.~\ref{fig:3}. Here we observe the feed-down from the $B_1^+/B_2^{*+}\to B^{*0}\pi^+$ and the $B_2^{*+}\to B^{0}\pi^+$ states. To improve fit convergence, the relative width between the $B_1^+$ and the $B_2^{*+}$ is fixed to 0.9 and the relative yield between the $B_2^{*+}\to B^{*0}\pi^+$ and $B_2^{*+}\to B^{0}\pi^+$ fixed to 0.93, from theoretical predictions \cite{spec-theo,bstth}.

The largest systematic uncertainty in the mean $Q$ values for the observed states, arise from variations in the selection requirements (0.95\%), and from the uncertainty on the $B^0$ mass. The mean values are corrected for possible biases, calculated from simulated toy samples generated from the experimental PDF. The final results are $M(B_1^+)=(5726.3\pm1.9\pm3.0\pm0.5)$\,MeV$/c^2$ and  $M(B_2^{*+})=(5739.0\pm3.3\pm1.6\pm0.3)$\,MeV$/c^2$, where the first uncertainty is statistical, the second systematical and the third one from the uncertainty on the $B^0$ meson mass. These masses are compatible with the masses of corresponding isospin partners $B_1^0$ and $B_2^{*0}$. In addition, we observe a signal significance of $9.9\sigma$ and $4.0\sigma$ for $B_1^+\to B^{*0}\pi^+$ and $B_2^{*+}\to B^{0}\pi^+$, respectively, corresponding to the first observation of these two decays. 

\subsection{Measurement of the $\Omega_b^-$ and $\Xi_b^-$ masses}
\label{masses} 
Using a  620\,pb$^{-1}$ integrated luminosity sample, we measure the masses of the strange $b$-baryons $\Omega_b^-$ and $\Xi_b^-$, which are constructed via the decay chains $J/\psi\Omega^-(\Lambda^0K^-)$ and $J/\psi\Xi^-(\Lambda^0\pi^-)$, respectively, with $J/\psi\to\mu^+\mu^-$ and $\Lambda^0\to p\pi^-$. The mass measurement of the $\Omega_b^-$ is of particular interest, since the masses reported for this state by the CDF and D$\emptyset$ collaborations are inconsistent at the 6$\sigma$ level \cite{omegamass}. Both decays share a similar topology and the presence of long-lived particles in the decay chain is exploited in the selection process. High background levels are observed near the interaction point, mainly from inclusive $J/\psi$ production, thus we reconstruct only candidates with lifetime above 0.3\,ps. The mass resolution of the strange $b$-hadrons is improved by applying a constraint to the mass of the daughters in the vertex fit. The final selection places restrictions on the particle identification and quality of the final state tracks. In addition, a momentum calibration is applied based on a large sample of $J/\psi\to\mu^+\mu^-$ candidates, that corrects for the description of the magnetic field and tracking system.

\begin{figure}
\resizebox{1.0\columnwidth}{!}{\includegraphics{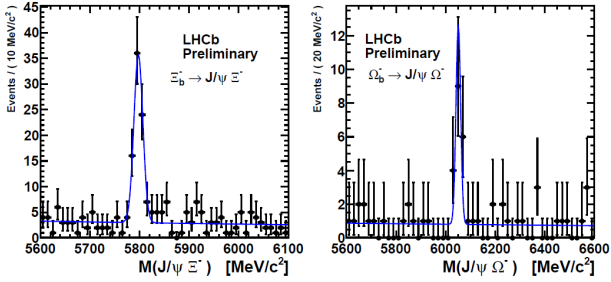} }
\caption{The invariant mass distributions for the selected $\Xi_b^-$ (left) and $\Omega_b^-$ (right) candidates. The fit projection is overlaid.}
\label{fig:4}      
\end{figure}

The invariant mass distributions for the selected $\Omega_b^-$ and $\Xi_b^-$ candidates is shown in Fig.~\ref{fig:4}. Signal candidates are described by Gaussian distributions, where the width for the $\Xi_b^-$ is extracted from simulated data and for the $\Omega_b^-$ it is estimated by scaling the $\Xi_b^-$ resolution by the ratio of the $\Omega_b^-$ and $\Xi_b^-$ masses. 

The systematic uncertainties in the measurement of the masses are dominated by the momentum scale calibration. A detailed description of the evaluation of this effect, can be found in Ref.~\cite{calib}. The final result for the masses are $M(\Xi_b^-) = 5796.5\pm1.2\pm1.2$\,MeV$/c^{2}$ and $M(\Omega_b^-)=6050.3\pm4.5\pm2.2$\,MeV$/c^{2}$, where the first uncertainty is statistical and the second one due to systematic effects. These results correspond are the best measurements of these masses to date. The measured $\Omega_b$ mass is compatible with the CDF measurement $M(\Omega_b^-) = 6054.4\pm6.9$\,MeV$/c^{2}$, but enlarge the discrepancy of the global average with the measurement performed by the D$\emptyset$ collaboration, $M(\Omega_b^-) = 6165\pm16$\,MeV$/c^{2}$.

\section{Conclusions}
\label{conclusions} 
We summarized a few selected results on heavy flavor production and spectroscopy at the LHCb detector. We expect many new results with the analysis of the 2011 dataset. 
The LHCb experiment has shown outstanding capabilities and is in a good position to explore the production mechanisms and spectra of states, as well as to produce competitive results in the heavy flavors sector.

\end{document}